\documentclass[preprint,3p,twocolumn,times,12pt]{elsarticle}

\usepackage{amssymb}
\usepackage{amsmath}
\usepackage{mathtools, cuted}
\usepackage{lineno}
\usepackage{color}
\usepackage[separate-uncertainty=true]{siunitx}
\usepackage{graphicx}

\journal{Nuclear Instr. and Meth. A}

\begin{document}

\begin{frontmatter}



\title{A Highly Segmented Neutron Polarimeter for A1}
\author[jgu]{R.~Spreckels}
\author[jgu]{M.~Hoek\corref{cor1}}
\author[jgu]{U.~Müller}
\author[jgu]{M.~Thiel}

\address[jgu]{Institut für Kernphysik, Johannes Gutenberg-Universität Mainz, D-55099 Mainz, Germany}

\cortext[cor1]{Email: hoek@uni-mainz.de}

\begin{abstract}
A new neutron polarimeter for measuring the neutron's electric form factor was designed and constructed to complement the A1 spectrometer setup at the Mainz Microtron (MAMI). The design is based on a previous polarimeter with significant improvements to halve the error of the extracted form factor.
A higher granularity of the polarimeter sections and a deeper first section on the one hand, and a faster readout employing Time-over-Threshold methods to measure the signal amplitudes combined with a high-precision FPGA-based TDC on the other hand will allow to achieve this goal. The performance of the new polarimeter during a first measurement campaign in 2019 using liquid hydrogen and deuterium targets will be discussed.
\end{abstract}

\begin{keyword}
Neutron polarimeter \sep Plastic Scintillator \sep FPGA TDC \sep Time-over-Threshold


\end{keyword}

\end{frontmatter}


\section{Introduction}
Measuring the neutron's electric form factor $G_{E,n}$ requires a sophisticated setup. The detection of neutrons with high momenta (several hundred \si{\mega\electronvolt\per c}) is notoriously difficult and needs large detector volumes.
This leads to a susceptibility for unwanted background processes, especially at high beam currents needed for such a measurement.

The general design concept and measurement principle, which was already used at A1 \cite{Ost99,Gen04}, is explained in Sec.~\ref{sec:concept}. The polarimeter's construction and readout and trigger concept are discussed in Sec.~\ref{sec:configuration} to \ref{sec:trigger}. Finally, the performance of the new polarimeter during a beam time with liquid hydrogen and deuterium targets in 2019 is shown in Sec.~\ref{sec:performance}.




\section{Concept and Polarimetry}
\label{sec:concept}
In order to measure the neutron electric form factor, $G_{E,n}$, in a model-independent way at the high $Q^2$, the same method that previously has been established at MAMI~\cite{Ost99,Gen04} will be used. 
A beam of polarised electrons hits an unpolarised, liquid deuterium target. In the quasielastic reaction $D(\vec{e} ,e' \vec{n}) p$ polarised neutrons are produced. The components of the outgoing neutron polarisation carry the information on its electric and magnetic form factors. For free (e,n) scattering Arnold, Carlson and Gross obtained \cite{Arn81}: 
\small
\begin{align} \label{polkomp}
{\cal P}_x & = -h P_e \frac{2 \sqrt{\tau(1+\tau)} \tan \frac{\vartheta_e}{2} G_E G_M}
{G_E^2 + \tau G_M^2 (1+2(1+\tau) \tan^2 \frac{\vartheta_e}{2})} \nonumber \\
{\cal P}_y & =  0 \\
{\cal P}_z & = h P_e \frac{2 \tau \sqrt{1+\tau+(1+\tau)^2 \tan^2 \frac{\vartheta_e}{2}} 
\tan \frac{\vartheta_e}{2} G_M^2}{G_E^2 + \tau G_M^2(1+2(1+\tau) \tan^2 
\frac{\vartheta_e}{2})} \nonumber
\end{align}
\normalsize

The coordinate frame $(x,y,z)$ is defined relative to the electron scattering plane and the direction of the momentum transfer:
$\hat{z}=\vec{q}/|\vec{q}|$,
$\hat{y} = \vec{p}_{e_i} \times \vec{p}_{e_f}/|\vec{p}_{e_i} \times
\vec{p}_{e_f}| $, $\hat{x} = \hat{y} \times \hat{z} $. The variable
$\tau=Q^2/4 M_n^2$ measures the negative squared 4-momentum transfer, $Q^2$, in units of the neutron mass, $M_n$. $P_e$ denotes the degree of longitudinal electron beam polarisation, and $h=\pm 1$ the electron
helicity. For quasielastic scattering off a neutron bound in deuterium, Eqs. \ref{polkomp} are basically maintained \cite{Are88} in the case of parallel kinematics, where the neutron is detected along the direction of the momentum transfer, $\vec{q}$. The ratio
\begin{equation} \label{polverh}
\frac{{\cal P}_x}{{\cal P}_z} = - \frac{1}{\sqrt{\tau + \tau(1+\tau) 
\tan^2 \frac{\vartheta_e}{2}}} \cdot \frac{G_{E,n}}{G_{M,n}} 
\end{equation}
is linear in $G_{E,n}$, and independent of the absolute value of $P_e$. The aim of the experiment is to measure this ratio.

\begin{figure}[h!]
	\centering
	\includegraphics[width=0.95\linewidth]{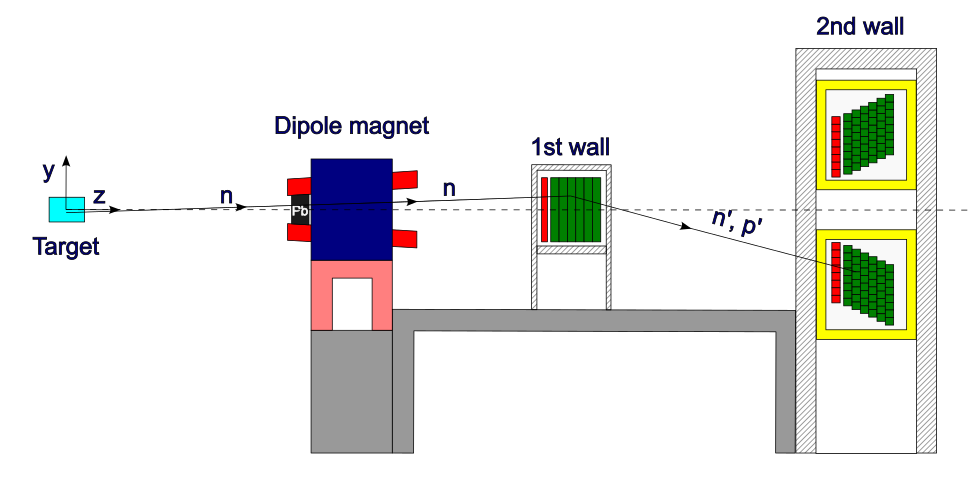}
	\caption{Sketch of Neutron Polarimeter setup with dipole magnet (not to scale). The scintillator (veto) bars in both walls are indicated in green (red).}
	\label{fig:ndet_sketch}
\end{figure}

The overall setup of the experiment is sketched in Fig.~\ref{fig:ndet_sketch}.
In order to tag the exclusive $(\vec{e},e'\vec{n})$ reaction, the scattered electrons and neutrons are detected in coincidence.
Energy and angles of the scattered electrons are measured with high resolution in one magnetic spectrometer of the A1 setup \cite{Blo98}. This guarantees clean separation of $\pi$ production events. An overdetermined reconstruction of the $D(\vec{e},e'\vec{n})p$ kinematics is achieved through the
additional measurement of the time-of-flight of the neutrons and their hit position in a segmented scintillator, first wall in Fig,~\ref{fig:ndet_sketch}, some \SI{6}{\meter} away from the target.
The neutron polarisation can be analysed in the detector itself \cite{Tad85}. This requires to detect the neutrons a second time in a split wall, second wall in Fig.~\ref{fig:ndet_sketch}, the two parts of which are positioned above and below the target (electron scattering plane).

The angular distribution of the scattered neutrons follows 
\small
\begin{align} \label{rate}
N^h(\Theta_n', \Phi_n') &= {\cal L}^h \varepsilon(\Theta_n', \Phi_n')
\sigma_0 [ 1+ h P_e {\cal A}_{\mathrm{eff}}(\Theta_n') \nonumber \\
&\qquad (P_x \sin \Phi_n'
+ P_y \cos \Phi_n')] \, ,
\end{align}
\normalsize
where, $\varepsilon(\Theta_n', \Phi_n')$ is the detection efficiency of the second wall scintillators, as a function of the polar and azimuthal angles of the first wall scattering. ${\cal L}^h$ denotes the helicity specific luminosity and $\sigma_0$ the polarisation-independent scattering cross-section. The effective analysing power, ${\cal A}_{\mathrm{eff}}$, is a priori unknown because
different reaction channels contribute in the scintillator material.

Taking advantage of the fact that a reversal of the beam helicity $h$ is equivalent to a shift of the azimuthal angle by \ang{180}, the asymmetry

\footnotesize
\begin{align} \label{asym}
A & = \frac{\sqrt{N^+(\Phi_n')N^-(\Phi_n'+180^\circ)} - \sqrt{N^+(\Phi_n'+180^\circ)N^-
(\Phi_n')}}{\sqrt{N^+(\Phi_n')N^-(\Phi_n'+180^\circ)} + \sqrt{N^+(\Phi_n'+180^\circ)N^-
(\Phi_n')}} \nonumber \\ 
& = P_e({\cal A}_{\mathrm{eff}}^t P_x \sin \Phi_n'
        + {\cal A}_{\mathrm{eff}}^n P_y \cos \Phi_n')
\end{align}
\normalsize
can be constructed in a way independent of ${\cal L}^h, \varepsilon$,
and $\sigma_0$. Here $N^\pm(\Phi_n')$ is the count rate in the second wall of scintillators. 

It has been demonstrated that the remaining problem of analysing power calibration of the polarimeter can be avoided by rotating the direction of the neutron spin in an external magnetic field \cite{Ost99}. In case of a field perpendicular to the neutron trajectory, the precession angle is proportional to the integrated
field strength,
\begin{align} \label{chi}
	\chi &= \frac{2 \mu_n}{\hbar c} \cdot \beta_n^{-1} \int B\; dl \nonumber \\
	&= \frac{-\ang{35.02}}{\si{\tesla\meter}}\cdot \beta_n^{-1} \int B\; dl \quad.
\end{align}
The same dipole magnet as in previous A3 \cite{Ost99} and A1 \cite{Gen04} experiments at MAMI is used. Since the $y$-component of the magnetic field is absolutely dominant, the spins precess in the $x$-$z$ plane.
The resulting transverse neutron polarisation behind the magnet is given by the superposition
\begin{equation}\label{prot}
P_\bot(\chi) = P_x \cos \chi - P_z \sin \chi =: P_0 \sin (\chi - \chi_0)
\end{equation}
with
\begin{equation}
P_0 = \sqrt{{P_x}^2+{P_z}^2} \quad.
\end{equation}
The transverse polarisation component, $P_\bot(\chi)$, varies with the
magnetic field strength, and thus the corresponding asymmetry
$A_\bot(\chi)$ in the second detector wall. For the particular case of
vanishing transverse polarisation, $P_\bot(\chi_0)=0$, after spin
precession by the angle $\chi_0$, the neutron electric form factor can
be experimentally determined from
\begin{align}\label{tanchi0}
	\tan \chi_0 &= \frac{A_x}{A_z} = \frac{P_e {\cal A}_{\mathrm{eff}} P_x}{P_e {\cal A}_{\rm eff} P_z} \nonumber \\
	&= - \frac{1}{\sqrt{\tau+\tau(1+\tau) \tan^2 \vartheta_e/2}} \cdot \frac{G_{E,n}}{G_{M,n}} \quad .
\end{align}
In the asymmetry ratio, both $P_e$ and ${\cal A}_{\rm eff}$ cancel out.
Their absolute values affect only the statistical error of the
determination of the zero crossing point $\chi_0$, but not its
absolute value. The electron beam polarisation has to be monitored,
in order to detect possible changes with time. This is done with a
Møller polarimeter~\cite{Bar01}.

\section{Detector Configuration}
\label{sec:configuration}
The neutron polarimeter consists of a total of 287 scintillator bars (cf. Tab.~\ref{tab:config}) in four different sizes (cf. Tab.~\ref{tab:bar_properties}), which are, as discussed in the previous section, arranged in two walls (cf. Fig.~\ref{fig:ndet_sketch}).
The sequence of scintillator types is the same in both walls: first a layer of thin veto bars and then a matrix of thicker signal bars.
The task of the veto bars is to detect charged particles entering the respective wall. Therefore, they are kept thin to minimise neutron interactions.
\begin{table}[!h]
	\caption{Overview of scintillator and PMT quantities for the new polarimeter design.}
	\label{tab:config}
	\centering
	\begin{tabular}{|c|c|c|c|}
		\hline
		& Veto & Signal & PMTs \\
		\hline
		first wall & 25 & 150 & 350 \\
		\hline
		second wall & 16 & 96 & 112 \\
		\hline
		\hline
		Total & 41 & 246 & 462 \\
		\hline
	\end{tabular}
\end{table}

Compared to the previous design \cite{seimetz_diplom, sanner_diplom, seimetz_phd, glazier_phd} (cf. Tab.~\ref{tab:config_old}), the new polarimeter has a deeper first wall and a finer segmentation of both walls. This will provide a better position resolution and reduce the rates in individual scintillators. 
\begin{table}[!h]
	\caption{Scintillator quantities for the old polarimeter design \cite{seimetz_diplom, sanner_diplom, seimetz_phd, glazier_phd}.}
	\label{tab:config_old}
	\centering
	\begin{tabular}{|c|c|c|}
		\hline
		& Veto & Signal \\
		\hline
		first wall & 15 & 30 \\
		\hline
		second wall & 8 & 24 \\
		\hline
		\hline
		Total & 23 & 54 \\
		\hline
	\end{tabular}
\end{table}

Except for the veto bars of the first wall, all bars were made of Eljen Technology\footnote{ELJEN TECHNOLOGY, 1300 W. Broadway, Sweetwater, TX 79556, USA} EJ-200 because of its long optical attenuation length and fast timing properties.
For the veto bars in the first wall, EJ-212 was chosen due to its superior light yield, as this type of bar has a large length-to-thickness ratio and its signal is important to reduce charged background. The second wall veto bars are less important for background reduction and thus are made from EJ-200 for economical reasons.
An overview of the different bar dimensions and used scintillator material is given in Tab.~\ref{tab:bar_properties}.
\begin{table}[!h]
	\caption{Bar dimensions and scintillator material for all for types used in the neutron polarimeter.}
	\label{tab:bar_properties}
	\centering
	\begin{tabular}{|p{1cm}|>{\centering}p{1.1cm}|p{1.1cm}|p{1.1cm}|p{1.2cm}|}  
		\hline
		Bar Type & Length [\si{\milli\meter}] & Width [\si{\milli\meter}] & Thick\-ness [\si{\milli\meter}] & Mate\-rial \\
		\hline
		\multicolumn{5}{|c|}{\emph{First Wall}}\\
		\hline
		Veto & 810 & 45 & 10 & EJ-212 \\
		\hline
		Signal & 800 & 30 & 45 & EJ-200 \\
		\hline
		\multicolumn{5}{|c|}{\emph{Second Wall}}\\
		\hline
		Veto & 1720 & 110 & 12 & EJ-200 \\
		\hline
		Signal & 1700 & 100 & 100 & EJ-200 \\
		\hline
	\end{tabular}
\end{table}

Every bar is at both ends connected via a tapered light guide to a photomultiplier tube (PMT). The light guides are glued to the scintillator bars using optical cement\footnote{Saint-Gobain BC-600}. Each bar is then wrapped with a thin aluminised mylar film to enhance the light yield and covered with black foil to shield from ambient light.

The PMTs are coupled to the light guides using a silicon elastomer\footnote{Elastosil RT 601} to allow for an easy replacement. Special attention was paid to the coaxial alignment of the PMTs of the first wall, as the tolerances for mounting were tight. 

For signal and veto bars of the first wall, the fast 10-stage PMT 9142SB from ET Enterprises\footnote{ET Enterprises, Ltd., 45 Riverside Way, Uxbridge, UB8 2YF, UK} with a diameter of \SI{29}{\milli\meter} with the C637ASN-05 passive voltage divider is chosen. 
All signal bars of the second wall are equipped with ET Enterprises 9822B PMTs with a diameter of \SI{78}{\milli\meter}, together with the passive voltage divider C638SFN-01. The veto bars are read out with Hamamatsu\footnote{Hamamatsu Photonics K.K., 325-6, Sunayama-cho, Naka-ku, Hamamatsu City, Shizuoka Pref., 430-8587, Japan} R580-17 PMTs with the passive voltage divider E2183-500MOD1.
 
All PMTs are shielded with $\mu$-metal tubes from the dipole magnet fringe field.

\begin{figure}[!h]
	\centering
	\includegraphics[height=0.9\linewidth,angle=270]{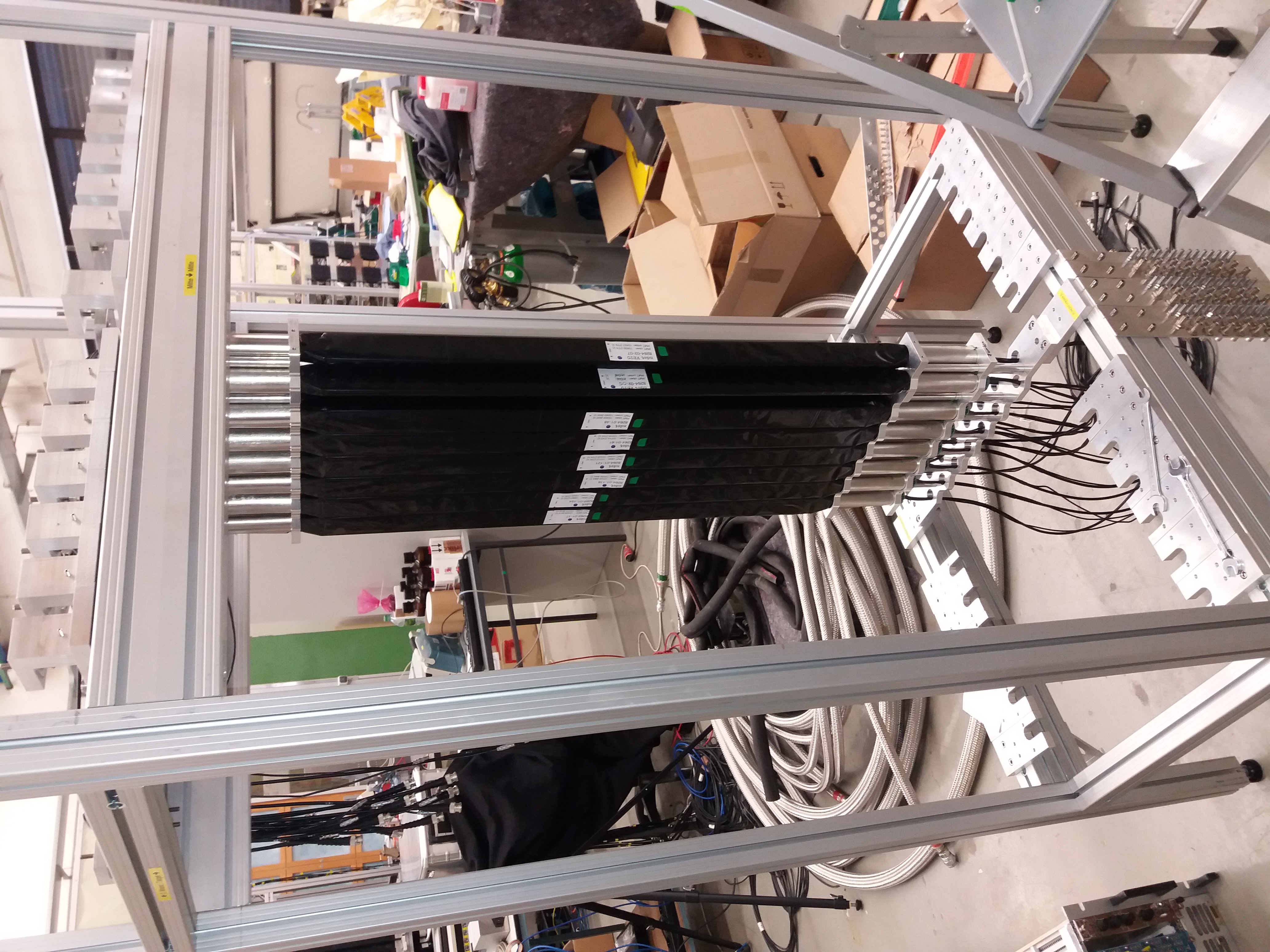}
	\caption{Image of a first wall module installed in the support frame during assembly.}
	\label{fig:firstmodule}
\end{figure}

The 150 signal bars of the first wall are arranged vertically in a matrix of 25$\times$6 bars (width $\times$ depth) with a veto bar in front of each of the 25 columns. Every second column is shifted back by half a scintillator thickness to allow for enough space for the PMTs. The entire first wall is composed of 13 special support structures or modules (see Fig.~\ref{fig:firstmodule}). 12 of these modules contain two columns of six signal and one veto bar each and one module with only one such column.

The second wall is placed \SI{2.5}{\meter} behind the first wall. The second wall consists of two frames (cf. Fig.~\ref{fig:ndet_sketch}) which are located one above the other with a gap of \SI{80}{\centi\meter} around beam level. Each frame consists of 48 signal and eight veto bars arranged horizontally on a stair-like support with an incline of \ang{18} in eight staggered rows.

\section{Readout}
\label{sec:readout}
The neutron polarimeter has to determine the time-of-flight of the neutrons and provide a precise location for any interaction within the detector volume to reconstruct the scattering angle. 
The interaction point is reconstructed by recording the arrival time $t_{1,2}$ of the scintillation light on either end of a scintillator bar.
The distance $d$ from the centre of a bar is then given by
\begin{equation}
	d = \frac{t_1 - t_2}{2} v_{eff}
\end{equation}
with $v_{eff}$ being the effective speed of light in a scintillator bar.

Moreover, information on the deposited energy will help to distinguish different particle species.

The readout electronics for the neutron polarimeter thus must provide a precise time-pickoff and amplitude measurement while coping with high rates (few \si{\mega\hertz}). Therefore, the readout system is based on the simplest time-pickoff method, leading-edge triggering, and time-over-threshold for amplitude determination which requires no additional delays in contrast to other methods. However, a very precise and fast TDC\footnote{ Time-to-Digital Converter} is needed for such a concept to work. Especially, since plastic scintillator signals have a fast rise time and short width. 

The frontend electronics is based on the NINO chip \cite{nino} developed by CERN. It provides eight input channels with very precise discriminators, a high-rate capability, and time-over-threshold encoding in the differential output signal. The NINO is operated in its single-ended mode. In addition, a programmable attenuator is placed before the NINO input to allow  adjustments for each channel to avoid operation outwith the NINO specifications. 
The frontend board was developed in-house and incorporates four NINO chips to provide 32 input channels. It has two operating modes: a local mode where an FPGA\footnote{Field Programmable Gate Array} add-on board controls all settings and receives the NINO output signals, and a remote mode where all settings are controlled via an I$^2$C connection from a computer.

The NINO output signals are then sent to multi-hit TDCs implemented in FPGAs on TRB3 boards \cite{trb3, trb3_01}. To avoid crosstalk by concurrent signals on adjacent lines, which could degrade time-pickoff, we use \SI{3}{\meter} long Cat.7 network cables with individually shielded twisted pairs and an additional shielding\footnote{Cat.7 S/FTP PIMF}.

The TRB3 TDC has a typical bin width of just \SI{15}{\pico\second} and provides 48 channels. Each channel has a ring buffer for storing hit information until a global trigger arrives. 
Each TRB3 board contains four such TDCs and a central FPGA for slow control management and data transport. Furthermore, the TRB TDC allows to implement certain user-defined trigger conditions. The use of a time window with respect to the global trigger allows to select only the relevant hits from the internal memory to reduce the required bandwidth. The data are sent from each TRB board in parallel to a readout PC, combined there and forwarded to the A1 DAQ system \cite{Dis01}.

In total, seven TRB3 boards are used to readout the neutron polarimeter: six boards in TDC configuration and one board as main trigger and slow control unit. Of the six TDC boards four are used to read out the first wall and two for the second wall.

For the actual measurements we expect large interaction rates at least in the first layers of the first wall. Therefore, the rate capability was investigated in a dedicated test experiment. A fully instrumented scintillator bar for the first wall was irradiated with an electron beam ($E_{beam} = \SI{855}{\mega\electronvolt}$). The beam intensity was varied by changing the Wehnelt voltage $U_W$ of the electron source. The count rate was measured using a scaler implemented on a Zynq FPGA board\footnote{Details here...}. Ten \SI{1}{\second}-intervals were measured and the average and standard deviation computed (s. Tab.~\ref{tab:rates}). A reference scintillation counter was installed behind the bar off the beam axis thus reducing the counting rates in this counter to avoid dead time effects. Then the ratio between the scintillator bar and the reference counter was computed and normalized to 1 at the lowest intensity. The error of this normalized ratio was computed according to
\begin{equation}
	\Delta R_{norm} = \sqrt{\left( \frac{\Delta R}{R_0} \right)^2 + \left( \frac{R \cdot \Delta R_0}{R_0^2} \right)^2}
\end{equation}
with R the raw ratio and $\Delta R$ its error, $R_0$ the raw ratio used for normalization and $\Delta R_0$ its error. 
This measurement was then repeated using the TRB3 readout to exclude any negative impact on the rate capability by this readout method. Here, the count rate could not be determined as precisely and the error was simply estimated as $\sqrt{N}$ (cf. Tab.~\ref{tab:rates}). 
\begin{table}
	\caption{Raw rates measured with Zynq and TRB3 readout. For the TRB measurement no error is given due to small systematic offsets.}
	\label{tab:rates}
	\small
	\begin{tabular}{|c|r|r|}
		\hline
		$U_W$ [\si{\volt}] & \multicolumn{1}{c|}{$\left\langle R\right\rangle _{Zynq}$ [\si{cps}]} & \multicolumn{1}{c|}{$R_{TRB}$ [\si{cps}]} \\
		\hline
		-13.0 & \num{74564(336)} & \num{82500(287)} \\
		\hline
		-12.9 & \num{110389(220)} & \num{119000(345)} \\
		\hline
		-12.8 & \num{162519(381)} & \num{175000(418)} \\
		\hline
		-12.7 & \num{238228(600)} & \num{260000(510)} \\
		\hline
		-12.6 & \num{351806(712)} & \num{390000(625)} \\
		\hline
		-12.5 & \num{511935(745)} & \num{556000(746)} \\
		\hline
		-12.4 & \num{751887(1131)} & \num{825000(908)} \\
		\hline
		-12.3 & \num{1089516(1384)} & \num{1190000(1091)} \\
		\hline
		-12.2 & \num{1567946(1366)} & \num{1750000(1323)} \\
		\hline
		-12.1 & \num{2216389(1561)} & \num{2500000(1581)} \\
		\hline
		-12.0 & \num{3066416(2076)} & \num{3550000(1884)} \\
		\hline
	\end{tabular}
	\normalsize
\end{table}

\begin{figure}[!h]
	\centering
	\includegraphics[width=0.95\linewidth]{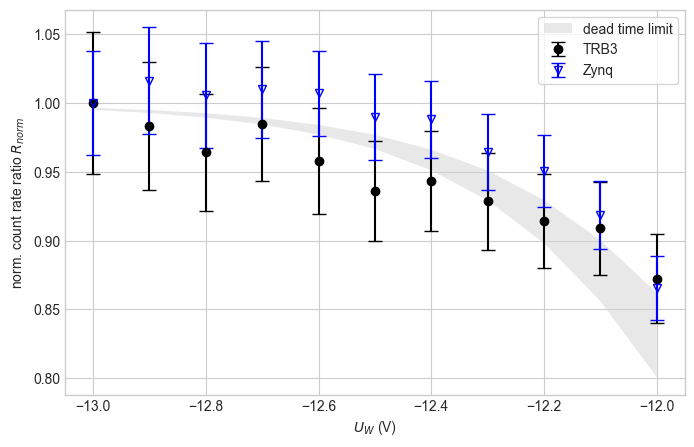}
	\caption{Test of the NDET readout rate capability. The normalized rate is shown as a function of the Wehnelt voltage (see text for explanations). The ratios are shown for both readout methods, Zynq and TRB3 board. The grey band indicates the expected ratio for signals with a Time-over-Threshold between \SI{45}{\nano\second} and \SI{65}{\nano\second} as observed during the experiment.}
	\label{fig:ratetestx1}
\end{figure}

The normalized ratios for both Zynq and TRB readout are shown in Fig.~\ref{fig:ratetestx1} together with the estimated limit due to dead time effects when assuming a signal width of \SIrange{45}{65}{\nano\second}. As can be seen, both readout method yield similar results close to the limit and thus are not introducing a bottleneck.

\section{Trigger}
\label{sec:trigger}
The neutron polarimeter must provide a trigger signal to the A1 trigger system to select coincidences with the spectrometer detecting the scattered electrons.

A valid neutron polarimeter event must contain at least one signal in each wall. A signal is defined as a simultaneously detected hit in both PMTs attached to a scintillator bar. The TRB3 TDC provides the capability to generate trigger signals based on its input signals. A list of channel pairs is written to the TDC configuration to form the basic coincidences at bar level. The coincidence time window is set to \SI{40}{\nano\second}. An OR is generated from all thus defined channel pairs and sent to the central FPGA on the TRB3 board. There, all four peripheral trigger signals are combined and sent out to the neutron polarimeter trigger logic. There, the signals of the first and second wall are combined first separately and then a coincidence of first and second wall triggers is formed. This final trigger signal is then forwarded to the global A1 trigger system.

The A1 trigger signal is then returned to the TRB3 system together with an event number. The trigger signal is forwarded to all TRB3 TDCs and the event number is included in the data sent to the A1 DAQ system.

The veto scintillator information is deliberately omitted from the neutron polarimeter trigger decision to avoid event losses due to poorly calibrated veto bars. As the individual bar trigger rate ($\approx \SI{1}{\mega\hertz}$) is well below the maximum sustainable rate ($\approx \SI{2.5}{\mega\hertz}$) at the maximum  beam current, this decision does not impact on the overall performance of the neutron polarimeter.

\section{Performance}
\label{sec:performance}
The new neutron polarimeter was installed at the A1  three-spectrometer setup at MAMI to measure the neutron's electric form factor at a momentum transfer of $Q^2 = 0.7\,(GeV/c)^2$ in quasi-free scattering of polarized electrons with a beam energy $E_e = \SI{855}{\mega\electronvolt}$ off a liquid deuterium target.

The initial calibration of the neutron polarimeter's scintillator bars was performed using cosmics and protons from a liquid hydrogen target and was used to adjust the high voltage of the PMTs to equalise the time-over-threshold spectra. When using the liquid hydrogen target, the elastically scattered electron was detected in coincidence by a spectrometer of the A1 setup.

\begin{figure}[!h]
	\begin{minipage}{0.49\linewidth}
		\centering
		\includegraphics[width=0.95\linewidth]{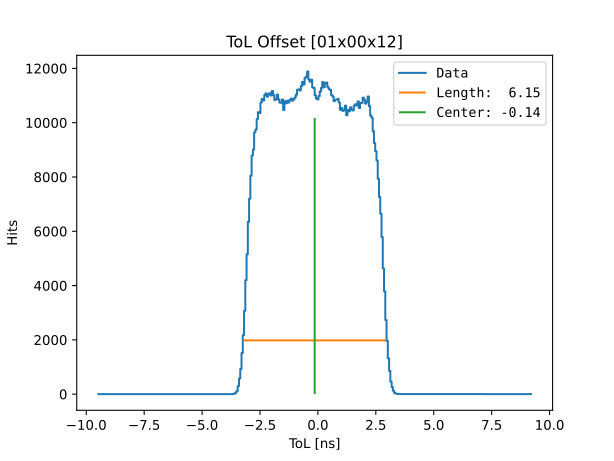}
		(a)
	\end{minipage}
	\hfill
	\begin{minipage}{0.49\linewidth}
		\centering
		\includegraphics[width=0.95\linewidth]{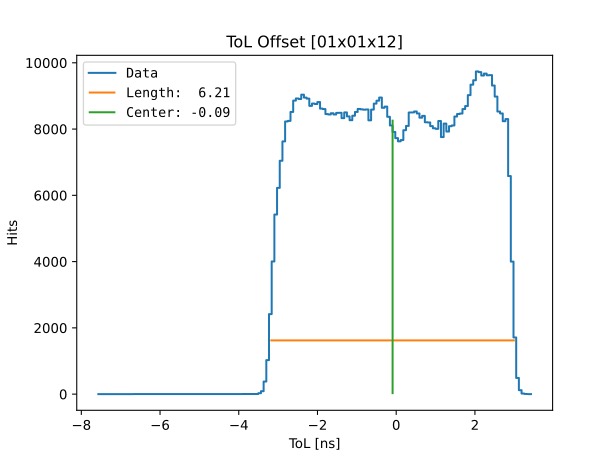}
		(b)
	\end{minipage}
	\hfil
	\begin{minipage}{0.49\linewidth}
		\centering
		\includegraphics[width=0.95\linewidth]{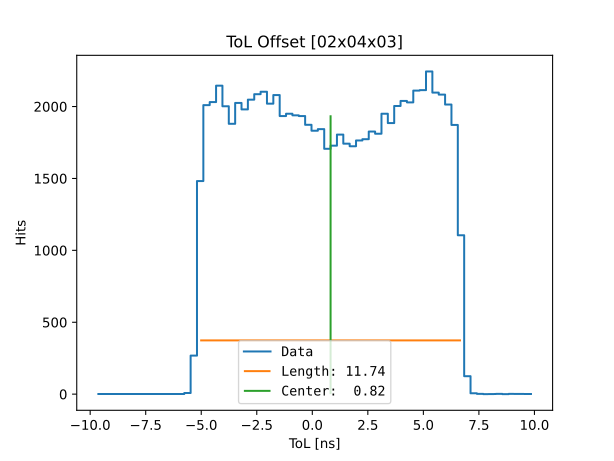}
		(c)
	\end{minipage}
	\caption{Time difference spectra for a veto (a) and a signal bar (b) in the first wall and across a signal bar in the second wall (c). Offset corrections (s. text) have already been applied.}
	\label{fig:barwidthtime}
\end{figure}

As a next step, the time difference spectra need to be corrected for any offsets introduced due to different signal propagation delays. 
The nearly homogeneous hit distribution across the bars in all detector elements is shown in Fig.~\ref{fig:barwidthtime}.
The left- and right-hand edge of the time difference distribution is determined and the corresponding offset computed.

An additional first wall signal bar was installed horizontally in front of the first wall at beam level to help with the position calibration. Requiring a hit in this additional bar allowed to validate the computed offsets for the first wall. 

In addition to the offsets, the width of these time difference distributions allows the determination of $v_{eff}$ for each type of bar. The height at which the width is measured is set to match a previous precise measurement of $v_{eff}$ for a first wall scintillator bar ($v_{eff} = \SI[per-mode=fraction]{138.4(0.28)}{\milli\meter\per\nano\second}$). The extracted values for $v_{eff}$ are given in Tab.~\ref{tab:effectice_speed}. Note that no value could extracted for the vetos of the second wall as their PMT response was not adjusted properly.
\begin{table}[!h]
	\centering
	\caption{Effective speed of light $v_{eff}$ for different detector elements.}
	\label{tab:effectice_speed}
	\begin{tabular}{|c|c|}
		\hline
		Detector element& $v_{eff}$   (\si[per-mode=fraction]{\milli\meter\per\nano\second}) \\
		\hline
		1$^{st}$ Wall Veto & \num{130.17(0.75)} \\
		\hline
		1$^{st}$ Wall Bar & \num{137.98(0.30)} \\
		\hline
		2$^{nd}$ Wall Bar & \num{139.64(0.27)} \\
		\hline
	\end{tabular} 
\end{table}

\begin{figure}[!h]
	\centering
	\includegraphics[width=0.95\linewidth]{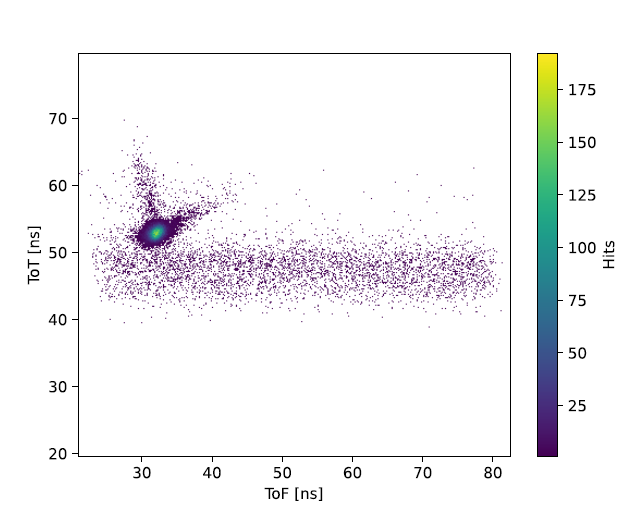}
	\caption{Average time-over-threshold (ToT) as function of Time-of-Flight (ToF) for a veto scintillator in the first wall using a hydrogen target.}
	\label{fig:veto1tot}
\end{figure}

Further steps in the calibration and analysis require information on the signal amplitude which is correlated to the time-over-threshold. A scintillator’s average  time-over-threshold $\left< T_{hit} \right>$ is defined as the geometric mean of both signals from its PMTs as
\begin{equation}
	\left< T_{hit} \right> = \sqrt{T_1 \cdot T_2}\quad .
\end{equation}
This additional information facilitates the separation of different particle types helping to suppress background contributions in the analysis. Fig.~\ref{fig:veto1tot} shows $\left< T_{hit} \right>$ of a veto bar of the first wall as a function of the time-of-flight. There is a clear correlation seen corresponding to protons scattered from the liquid hydrogen target and a distinct band of uncorrelated background at lower time-over-threshold values.

\begin{figure}[!h]
	\centering
	\begin{minipage}{0.49\linewidth}
		\centering
		\includegraphics[width=0.95\linewidth]{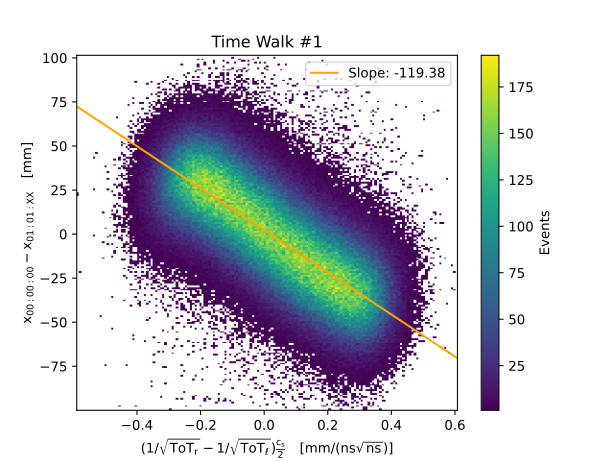}
		(a)
	\end{minipage}
	\hfill
	\begin{minipage}{0.49\linewidth}
		\centering
		\includegraphics[width=0.95\linewidth]{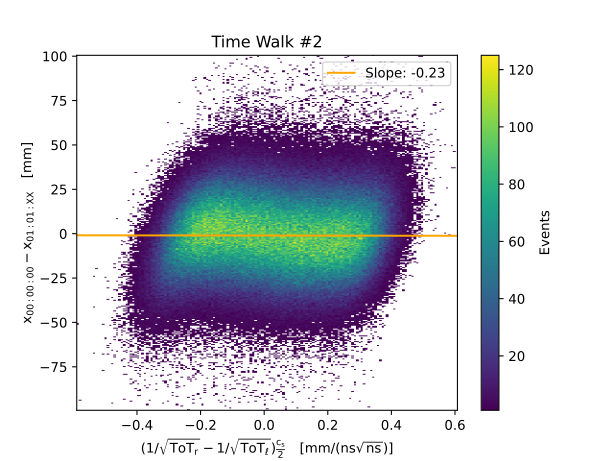}
		(b)
	\end{minipage}
	\caption{Deviation of reconstructed from true position along a signal bar before (a) and after (b) walk-corrections. For details, please see text above.}
	\label{fig:timewalkhorizontal}
\end{figure}

Furthermore, $\left< T_{hit} \right>$ can be used to improve the time-pickoff as leading edge discrimination suffers from time-walk effects. To study these effects, the additional horizontal bar was used. The difference of the reconstructed hit position $d$ and the geometrical position $d'$ of the first wall bars behind hit too can be expressed in terms of the measured time-over-threshold of each PMT signal as
\begin{equation}
	d - d' = m \left( \frac{1}{\sqrt{T_2}} - \frac{1}{\sqrt{T_1}} \right) \frac{v_{eff}}{2} \quad ,
\end{equation}
with the slope parameter $m$, assuming that the time-over-threshold is proportional to the signal amplitude and that the signal shape can be approximated by a second-order polynomial up to the set threshold level. Fig.~\ref{fig:timewalkhorizontal} clearly shows the time-walk behaviour and allows to determine the slope parameter $m = -\SI{119.16}{\nano\second\sqrt{\mathrm{\nano\second}}}$. After applying the corresponding corrections to the measured time-pickoff, this correlation has vanished. 
Since the horizontal bar has the same properties as the signal bars of the first wall, this value for $m$ is used to correct the time-pickoff for all signal bars in the first wall. However, such a measurement cannot not be done for other scintillator bars of the neutron polarimeter.

\begin{figure}[!h]
	\centering
	\includegraphics[width=0.95\linewidth]{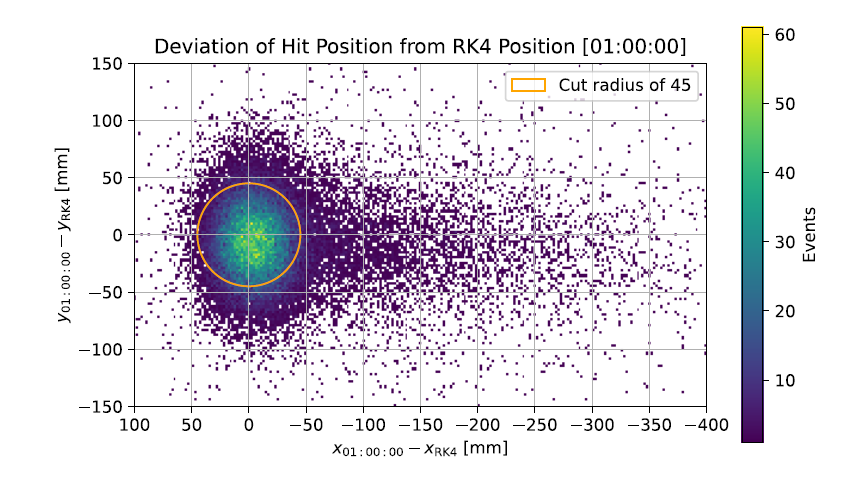}
	\caption{Difference between predicted and reconstructed hit position for protons in the first wall. Data was taken with a liquid hydrogen target.}
	\label{fig:hitposition}
\end{figure}

The position reconstruction in the first wall is cross checked using data with a liquid hydrogen target. Here, the initial direction and velocity of the proton can be calculated from the scattered electron kinematics. Propagating this track through the dipole magnet, a hit position on the first wall can be computed and compared to the reconstructed position (cf. Fig.~\ref{fig:hitposition}) showing an excellent agreement. Most hits are within a circle with a radius of \SI{45}{\milli\meter} around the predicted hit position, the tails caused by interactions with the target cell.

The events within the circled area are then used to correct time-of-flight offsets for each bar in the first wall. For the second wall, a different approach is required to determine these offsets as there is, by design, no direct line-of-sight to the target. Selecting light relativistic particles, that are scattered within in the first wall, allow to measure the time of flight between the two walls and extract the offset parameters.

For the final extraction of the neutron electric form factor, two different scattering reactions in the neutron polarimeter have to be distinguished: either the incoming neutron is scattered in the first wall and then detected in the second (\emph{nn}-channel) or the incoming neutron transfers in the first wall its momentum to a proton which is then detected in the second wall (\emph{np}-channel). 

The coarse separation of these two cases is based on the response of the veto bars and the hit multiplicity in each wall. However, it was found that the performance of the veto layers was insufficient and, therefore, the first layer of each wall was included in the veto decision.

For brevity, only the \emph{nn}-channel is discussed below as the detection of protons is by far easier. The pre-selection of these events requires no hits in the veto and first signal bar layer of both walls.

\begin{figure}[!h]
	\centering
	\includegraphics[width=0.95\linewidth]{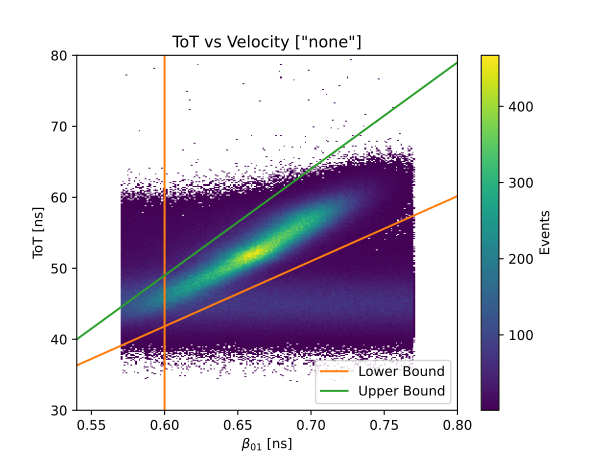}
	\caption{Correlation of time over threshold (ToT) with neutron velocity $\beta_{01}$ for \emph{nn}-channel in the first wall. Only events between the diagonal lines and with $\beta_{01}$ above \num{0.6} are selected for nn-channel analysis.}
	\label{fig:vel1-tot1nonenn}
\end{figure}

This selection can be further improved due to the excellent performance of the neutron polarimeter.
A clear correlation between the deposited energy $\left< T_{hit} \right>$ and the incoming neutron's velocity $\beta_{01}$ is visible, see Fig.~\ref{fig:vel1-tot1nonenn}. The corresponding cut limits are indicated by the coloured lines.

\begin{figure}[!h]
	\centering
	\includegraphics[width=0.95\linewidth]{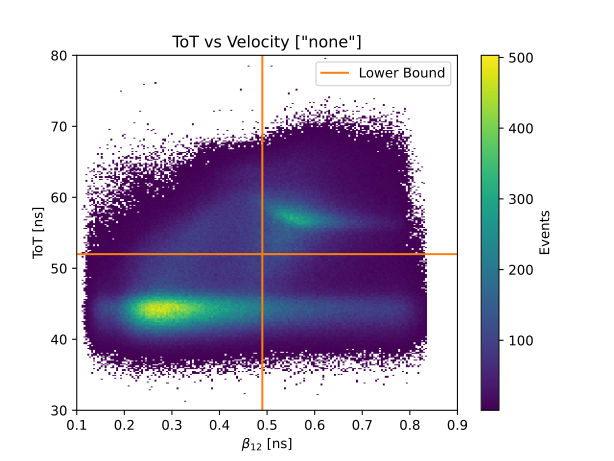}
	\caption{Correlation second wall ToF and ToT.
	Correlation of time over threshold (ToT) with neutron velocity $\beta_{12}$ for \emph{nn}-channel in the second wall. Only events in the upper right quarter are selected for the nn-channel analysis.}
	\label{fig:vel2-tot2nonenn}
\end{figure}

The same correlation can be plotted for the second wall (s. Fig.~\ref{fig:vel2-tot2nonenn}) with the additional requirement of no hit in the last layer of the first wall to improve the pre-selection. The distinction is less clear than for the first wall but a full analysis reveals that the relevant neutron events are in the upper right corner.

This improved selection of the relevant events in the neutron polarimeter allows to reduce the background, which would otherwise dilute the measured azimuthal asymmetry. 

\begin{figure}[!h]
	\centering
	\includegraphics[width=0.95\linewidth]{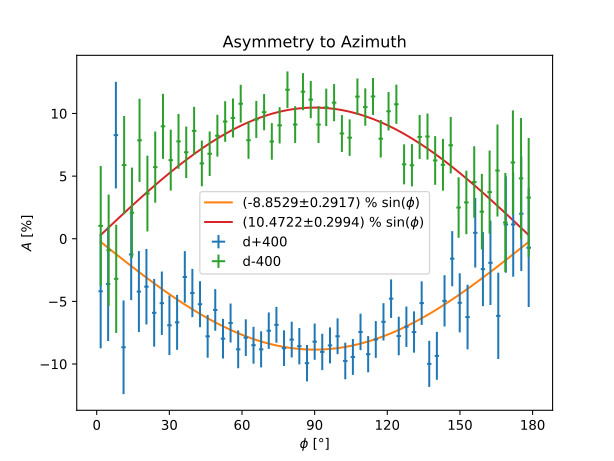}
	\caption{Final azimuthal asymmetry for full analysis with magnet currents of $\pm$\SI{400}{\ampere}.}
	\label{fig:asymmetry}
\end{figure}

With all calibrations and selection cuts in place, the azimuthal asymmetry (cf. Eq.~\ref{asym}) can be extracted for different magnetic field strengths. The asymmetry for the maximum magnet current is shown in Fig.~\ref{fig:asymmetry}.

\section{Conclusions}
\label{sec:conclusions}
The newly built neutron polarimeter for A1 spectrometer setup at MAMI includes an improved design for the detector layers with a larger depth and higher segmentation. Furthermore, the readout electronics is designed to cope with high rates in individual channels. To facilitate this approach, time-pickoff is achieved by a leading edge discriminator nad the signal amplitude is measured by a time over threshold approach avoiding unnecessary dead times due to digitization. A TRB3 system provides the required high-precision fast TDC to enable this scheme.

This neutron polarimeter was used in a three week long beam time in 2019 to measure the neutron's electric form factor at $Q^2 = 0.7\,(GeV/c)^2$. The calibration results shown above demonstrate the excellent performance of the new detector system and corroborate the validity of the design decisions.
\section*{Acknowledgements}
\label{sec:ack}
We acknowledge the MAMI accelerator group and all
the workshop staff members for the outstanding support.
Furthermore, the authors would like to thank the TRB3 collaboration for their outstanding support to set up the readout system for the neutron polarimeter.

Funding: This work was supported by the Deutsche Forschungsgemeinschaft through the Collaborative Research Center 1044 and grant no INST 247/771-1 FUGG and the Federal
State of Rhineland-Palatinate.

\end{document}